\documentclass[10pt,english]{article}
\usepackage{ae,aecompl}
\usepackage[T1]{fontenc}
\usepackage[latin9]{inputenc}
\usepackage{color}
\usepackage{babel}
\usepackage{array}
\usepackage{verbatim}
\usepackage{bm}
\usepackage{multirow}
\usepackage{amsmath}
\usepackage{amssymb}
\usepackage{graphicx}
\usepackage{esint}
\usepackage[unicode=true,pdfusetitle,
 bookmarks=true,bookmarksnumbered=false,bookmarksopen=false,
 breaklinks=false,pdfborder={0 0 1},backref=false,colorlinks=false]
 {hyperref}

\makeatletter

\providecommand{\tabularnewline}{\\}

\usepackage{babel}
\usepackage[font=small,labelfont=bf]{caption}
\@ifundefined{definecolor}{\@ifundefined{definecolor}
 {\usepackage{color}}{}
}{}
\@ifundefined{definecolor}{\@ifundefined{definecolor}
 {\usepackage{color}}{}
}{}
\@ifundefined{definecolor}{\@ifundefined{definecolor}
 {\@ifundefined{definecolor}
 {\@ifundefined{definecolor}
 {\usepackage{color}}{}
}{}
}{}
}{}
\@ifundefined{definecolor}{\@ifundefined{definecolor}
 {\@ifundefined{definecolor}
 {\@ifundefined{definecolor}
 {\@ifundefined{definecolor}
 {\usepackage{color}}{}
}{}
}{}
}{}
}{}
\@ifundefined{definecolor}{\@ifundefined{definecolor}
 {\@ifundefined{definecolor}
 {\@ifundefined{definecolor}
 {\@ifundefined{definecolor}
 {\@ifundefined{definecolor}
 {\usepackage{color}}{}
}{}
}{}
}{}
}{}
}{}
\@ifundefined{definecolor}{\@ifundefined{definecolor}
 {\@ifundefined{definecolor}
 {\@ifundefined{definecolor}
 {\@ifundefined{definecolor}
 {\@ifundefined{definecolor}
 {\@ifundefined{definecolor}
 {\usepackage{color}}{}
}{}
}{}
}{}
}{}
}{}
}{}
\@ifundefined{definecolor}{\@ifundefined{definecolor}
 {\@ifundefined{definecolor}
 {\@ifundefined{definecolor}
 {\@ifundefined{definecolor}
 {\@ifundefined{definecolor}
 {\@ifundefined{definecolor}
 {\@ifundefined{definecolor}
 {\usepackage{color}}{}
}{}
}{}
}{}
}{}
}{}
}{}
}{}
\@ifundefined{definecolor}{\@ifundefined{definecolor}
 {\@ifundefined{definecolor}
 {\@ifundefined{definecolor}
 {\@ifundefined{definecolor}
 {\@ifundefined{definecolor}
 {\@ifundefined{definecolor}
 {\@ifundefined{definecolor}
 {\@ifundefined{definecolor}
 {\usepackage{color}}{}
}{}
}{}
}{}
}{}
}{}
}{}
}{}
}{}
\@ifundefined{definecolor}{\@ifundefined{definecolor}
 {\@ifundefined{definecolor}
 {\@ifundefined{definecolor}
 {\@ifundefined{definecolor}
 {\@ifundefined{definecolor}
 {\@ifundefined{definecolor}
 {\@ifundefined{definecolor}
 {\@ifundefined{definecolor}
 {\@ifundefined{definecolor}
 {\usepackage{color}}{}
}{}
}{}
}{}
}{}
}{}
}{}
}{}
}{}
}{}
\@ifundefined{definecolor}{\@ifundefined{definecolor}
 {\@ifundefined{definecolor}
 {\@ifundefined{definecolor}
 {\@ifundefined{definecolor}
 {\@ifundefined{definecolor}
 {\@ifundefined{definecolor}
 {\@ifundefined{definecolor}
 {\@ifundefined{definecolor}
 {\@ifundefined{definecolor}
 {\@ifundefined{definecolor}
 {\usepackage{color}}{}
}{}
}{}
}{}
}{}
}{}
}{}
}{}
}{}
}{}
}{}
\@ifundefined{definecolor}{\@ifundefined{definecolor}
 {\@ifundefined{definecolor}
 {\@ifundefined{definecolor}
 {\@ifundefined{definecolor}
 {\@ifundefined{definecolor}
 {\@ifundefined{definecolor}
 {\@ifundefined{definecolor}
 {\@ifundefined{definecolor}
 {\@ifundefined{definecolor}
 {\@ifundefined{definecolor}
 {\@ifundefined{definecolor}
 {\usepackage{color}}{}
}{}
}{}
}{}
}{}
}{}
}{}
}{}
}{}
}{}
}{}
}{}
\@ifundefined{definecolor}{\@ifundefined{definecolor}
 {\@ifundefined{definecolor}
 {\@ifundefined{definecolor}
 {\@ifundefined{definecolor}
 {\@ifundefined{definecolor}
 {\@ifundefined{definecolor}
 {\@ifundefined{definecolor}
 {\@ifundefined{definecolor}
 {\@ifundefined{definecolor}
 {\@ifundefined{definecolor}
 {\@ifundefined{definecolor}
 {\usepackage{color}}{}
}{}
}{}
}{}
}{}
}{}
}{}
}{}
}{}
}{}
}{}
}{}
\@ifundefined{definecolor}{\@ifundefined{definecolor}
 {\@ifundefined{definecolor}
 {\@ifundefined{definecolor}
 {\@ifundefined{definecolor}
 {\@ifundefined{definecolor}
 {\@ifundefined{definecolor}
 {\@ifundefined{definecolor}
 {\@ifundefined{definecolor}
 {\@ifundefined{definecolor}
 {\@ifundefined{definecolor}
 {\@ifundefined{definecolor}
 {\@ifundefined{definecolor}
 {\@ifundefined{definecolor}
 {\usepackage{color}}{}
}{}
}{}
}{}
}{}
}{}
}{}
}{}
}{}
}{}
}{}
}{}
}{}
}{}
\@ifundefined{definecolor}{\@ifundefined{definecolor}
 {\@ifundefined{definecolor}
 {\@ifundefined{definecolor}
 {\@ifundefined{definecolor}
 {\@ifundefined{definecolor}
 {\@ifundefined{definecolor}
 {\@ifundefined{definecolor}
 {\@ifundefined{definecolor}
 {\@ifundefined{definecolor}
 {\@ifundefined{definecolor}
 {\@ifundefined{definecolor}
 {\@ifundefined{definecolor}
 {\@ifundefined{definecolor}
 {\@ifundefined{definecolor}
 {\usepackage{color}}{}
}{}
}{}
}{}
}{}
}{}
}{}
}{}
}{}
}{}
}{}
}{}
}{}
}{}
}{}
\usepackage{bm}

\@ifundefined{definecolor}{\@ifundefined{definecolor}
 {\@ifundefined{definecolor}
 {\@ifundefined{definecolor}
 {\@ifundefined{definecolor}
 {\@ifundefined{definecolor}
 {\@ifundefined{definecolor}
 {\@ifundefined{definecolor}
 {\@ifundefined{definecolor}
 {\@ifundefined{definecolor}
 {\@ifundefined{definecolor}
 {\@ifundefined{definecolor}
 {\@ifundefined{definecolor}
 {\@ifundefined{definecolor}
 {\@ifundefined{definecolor}
 {\@ifundefined{definecolor}
 {\usepackage{color}}{}
}{}
}{}
}{}
}{}
}{}
}{}
}{}
}{}
}{}
}{}
}{}
}{}
}{}
}{}
}{}

\usepackage{bm}

\@ifundefined{definecolor}{\@ifundefined{definecolor}
 {\@ifundefined{definecolor}
 {\@ifundefined{definecolor}
 {\@ifundefined{definecolor}
 {\@ifundefined{definecolor}
 {\@ifundefined{definecolor}
 {\@ifundefined{definecolor}
 {\@ifundefined{definecolor}
 {\@ifundefined{definecolor}
 {\@ifundefined{definecolor}
 {\@ifundefined{definecolor}
 {\@ifundefined{definecolor}
 {\@ifundefined{definecolor}
 {\@ifundefined{definecolor}
 {\@ifundefined{definecolor}
 {\@ifundefined{definecolor}
 {\usepackage{color}}{}
}{}
}{}
}{}
}{}
}{}
}{}
}{}
}{}
}{}
}{}
}{}
}{}
}{}
}{}
}{}
}{}

\usepackage{amsfonts}

\usepackage{epsfig}

\usepackage{latexsym}

\def\lesssim{\mathrel{\hbox{\rlap{\hbox{\lower4pt\hbox{$\sim$}}}\hbox{$<$}}}}
\def\gtrsim{\mathrel{\hbox{\rlap{\hbox{\lower4pt\hbox{$\sim$}}}\hbox{$>$}}}}

\usepackage{aecompl}
\usepackage{cite}
\usepackage{amssymb}
\usepackage{pifont}
\date{}

\makeatother

\begin{document}
\title{\textbf{{Spinning cylinders in general relativity: a canonical form
for the Lewis metrics of the Weyl class}}}
\author{L. Filipe O. Costa$^{1}$\thanks{lfilipecosta@tecnico.ulisboa.pt},
José Natário$^{1}$\thanks{jnatar@math.ist.utl.pt}, and N. O. Santos$^{2}$\thanks{Nilton.Santos@obspm.fr}
\\
 \\
{\small{} }{\normalsize{}{\em $^{1}$CAMGSD -- Departamento de
Matemática, Instituto Superior Técnico,} }\\
{\normalsize{} {\em Universidade de Lisboa, 1049-001, Lisboa, Portugal}}\\
{\normalsize{} {\em $^{2}$Sorbonne Université, UPMC Université
Paris 06, LERMA, UMRS8112 CNRS,} }\\
{\normalsize{} {\em Observatoire de Paris-Meudon, 5, Place Jules
Janssen, F-92195 Meudon Cedex, France} }}
\maketitle
\begin{abstract}
In the main article {[}CQG \textbf{38} (2021) 055003{]}, a new \textquotedblleft canonical\textquotedblright{}
form for the Lewis metrics of the Weyl class has been obtained, depending
only on three parameters --- Komar mass and angular momentum per
unit length, plus the angle deficit --- corresponding to a coordinate
system fixed to the \textquotedblleft distant stars\textquotedblright{}
and an everywhere timelike Killing vector field. Such form evinces
the local but non-global static character of the spacetime, and striking
parallelisms with the electromagnetic analogue. We discuss here its
generality, main physical features and important limits (the Levi-Civita
static cylinder, and spinning cosmic strings). We contrast it on geometric
and physical grounds with the Kerr spacetime --- as an example of
a metric which is locally non-static.\\
 \\
 \textbf{Keywords:} gravitomagnetism;~1+3 quasi-Maxwell formalism;~Sagnac
effect; gravitomagnetic clock effect;~synchronization gap;~local
and global staticity;~Levi-Civita solution;~cosmic strings
\end{abstract}
\tableofcontents{}

\section{Introduction}

The general stationary solution of the vacuum Einstein field equations
with cylindrical symmetry are the Lewis metrics \cite{Lewis:1932,SantosGRG1995,SantosCQG1995,GriffithsPodolsky2009}
\begin{equation}
ds^{2}=-f(dt^{2}+Cd\varphi)^{2}+r^{(n^{2}-1)/2}(dr^{2}+dz^{2})+\frac{r^{2}}{f}d\varphi^{2}\ ;\label{eq:LewisMetric}
\end{equation}
\begin{equation}
f=ar^{1-n}-\frac{c^{2}r^{n+1}}{n^{2}a}\ ;\qquad C=\frac{cr^{n+1}}{naf}+b\ ,\label{eq:LewisFunctions}
\end{equation}
usually interpreted as describing the exterior gravitational field
produced by infinitely long rotating cylinders. They divide into two
classes: (i) the Weyl class, when all the constants $n$, $a$, $b$,
and $c$ are real; (ii) the Lewis class, for $n$ imaginary {[}implying
in turn $c$ real and $a$ and $b$ complex, in order for the line
element \eqref{eq:LewisMetric} to be real\cite{GriffithsPodolsky2009,SantosCQG1995}{]}.

In the main article, Ref. \cite{Cilindros}, we have shown that
the Weyl class metrics can be written in the ``canonical'' form
\[
ds^{2}=-\frac{r^{4\lambda_{{\rm m}}}}{\alpha}\left(dt-\frac{j}{\lambda_{{\rm m}}-1/4}d\phi\right)^{2}+r^{4\lambda_{{\rm m}}(2\lambda_{{\rm m}}-1)}(dr^{2}+dz^{2})+\alpha r^{2(1-2\lambda_{{\rm m}})}d\phi^{2}\ ,
\]
depending only on three parameters with a clear physical significance:
the Komar mass ($\lambda_{{\rm m}}$) and angular momentum ($j$)
per unit $z$-length, plus the parameter $\alpha$ governing the angle
deficit. This form allows for a transparent comparison with the Levi-Civita
non-rotating cylinder --- archetype of the contrast between local
and global staticity --- which Ref. \cite{Cilindros} discusses
in detail both on physical and geometrical grounds. Therein its matching
to the van Stockum \cite{Stockum1938} cylinder in star fixed coordinates
is also shown. Here we revise some main features of the solution,
focusing on its generality {[}and redundancies of the more usual form
\eqref{eq:LewisMetric}-\eqref{eq:LewisFunctions}{]}, notable limits,
and physical properties, with special attention to those less developed
in Ref. \cite{Cilindros}. Addressing the question posed to us
in the discussion following the presentation of this work at MG16\cite{TalkMG16},
we focus here on its comparison with a non-static (globally and locally)
stationary solution, exemplified by the Kerr spacetime.

\section{Stationary spacetimes and levels of gravitomagnetism\label{sec:Stationary-spacetimes-and-GM}}

The line element $ds^{2}=g_{\alpha\beta}dx^{\alpha}dx^{\beta}$ of
a stationary spacetime can generically be written as 
\begin{equation}
ds^{2}=-e^{2\Phi}(dt-\mathcal{A}_{i}dx^{i})^{2}+h_{ij}dx^{i}dx^{j}\ ,\label{eq:StatMetric}
\end{equation}
where $e^{2\Phi}=-g_{00}$, $\Phi\equiv\Phi(x^{j})$, $\mathcal{A}_{i}\equiv\mathcal{A}_{i}(x^{j})=-g_{0i}/g_{00}$,
and $h_{ij}\equiv h_{ij}(x^{k})=g_{ij}+e^{2\Phi}\mathcal{A}_{i}\mathcal{A}_{j}$.
Observers of 4-velocity $u^{\alpha}=e^{-\Phi}\partial_{t}^{\alpha}\equiv e^{-\Phi}\delta_{0}^{\alpha}$,
whose worldlines are tangent to the timelike Killing vector field
$\partial_{t}$, are \emph{at rest} in the coordinate system of (\ref{eq:StatMetric}).
They are dubbed ``static'' or ``laboratory'' observers. The quotient
of the spacetime by their worldlines yields a 3-D Riemannian manifold
$\Sigma$ with metric $h_{ij}$ (called the ``spatial metric''),
which measures the spatial distances between neighboring laboratory
observers \cite{LandauLifshitz}. It is identified in spacetime with
the \emph{space projector} with respect to $u^{\alpha}$, $h_{\alpha\beta}\equiv u_{\alpha}u_{\beta}+g_{\alpha\beta}$.
Let $U^{\alpha}=dx^{\alpha}/d\tau$ be the 4-velocity of a test point
particle in geodesic motion. The space components of the geodesic
equation $DU^{\alpha}/d\tau=0$ yield \cite{Cilindros,LandauLifshitz,ZonozBell1998,ManyFaces,Analogies}
\begin{equation}
\frac{\tilde{D}\vec{U}}{d\tau}=\gamma\left[\gamma\vec{G}+\vec{U}\times\vec{H}\right]\ ;\qquad\vec{G}=-\tilde{\nabla}\Phi\ ;\qquad\vec{H}=e^{\Phi}\tilde{\nabla}\times\vec{\mathcal{A}}\ ,\label{eq:QMGeo}
\end{equation}
where $\gamma=-U^{\alpha}u_{\alpha}$ is the Lorentz factor between
$U^{\alpha}$ and $u^{\alpha}$, $\tilde{\nabla}$ denotes covariant
differentiation with respect to the spatial metric $h_{ij}$ (i.e.,
the Levi-Civita connection of $\Sigma$, $\tilde{\nabla}_{j}X^{i}=X_{\ ,j}^{i}+\Gamma(h)_{jk}^{i}X^{k}$,
for some spatial vector $\vec{X}$), $\tilde{D}/d\tau\equiv U^{i}\tilde{\nabla}_{i}$,
so that $\tilde{D}\vec{U}/d\tau$ describes the acceleration of the
3-D curve obtained by projecting the time-like geodesic onto the space
manifold $\Sigma$, being $\vec{U}$ its tangent vector. The latter
is identified in spacetime with the projection of $U^{\alpha}$ onto
$\Sigma$: $(\vec{U})^{\alpha}=h_{\ \beta}^{\alpha}U^{\beta}$ {[}so
its space components equal those of $U^{\alpha}$, $(\vec{U})^{i}=U^{i}${]}.
The spatial vectors $\vec{G}$ and $\vec{H}$ (living on $\Sigma$)
are dubbed, respectively, ``gravitoelectric'' and ``gravitomagnetic''
fields {[}or, jointly, ``gravitoelectromagnetic'' (GEM) fields{]}.
These play in Eq. (\ref{eq:QMGeo}) roles analogous to those of the
electric ($\vec{E}$) and magnetic ($\vec{B}$) fields in the Lorentz
force equation, $DU^{i}/d\tau=(q/m)[\gamma\vec{E}+\vec{U}\times\vec{B}]^{i}$,
and are identified in spacetime, respectively, with minus the acceleration
and twice the vorticity of the laboratory observers: 
\begin{equation}
G^{\alpha}=-\nabla_{\mathbf{u}}u^{\alpha}\equiv-u_{\ ;\beta}^{\alpha}u^{\beta}\;;\qquad H^{\alpha}=\epsilon^{\alpha\beta\gamma\delta}u_{\gamma;\beta}u_{\delta}\;.\label{eq:GEM Fields Cov}
\end{equation}
They motivate also dubbing the scalar $\Phi$ and the vector $\vec{\mathcal{\mathcal{A}}}$
gravitoelectric and gravitomagnetic potentials, respectively. 

Other realizations of the analogy arise in the equations of motion
for a ``gyroscope'' (i.e., a spinning pole-dipole particle) in a
gravitational field and a magnetic dipole in a electromagnetic field.
According to the Mathisson-Papapetrou equations \cite{Mathisson:1937zz,Papapetrou:1951pa},
under the Mathisson-Pirani spin condition \cite{Costa:2012cy}, the
spin vector $S^{\alpha}$ of a gyroscope of 4-velocity $U^{\alpha}$
is Fermi-Walker transported along its center of mass worldline, $DS^{\alpha}/d\tau=S^{\mu}a_{\mu}U^{\alpha}$,
where $a^{\alpha}\equiv DU^{\alpha}/d\tau$. If the gyroscope's center
of mass is at rest in the coordinate system of (\ref{eq:StatMetric})
($U^{\alpha}=u^{\alpha}$) the space part of this equation yields\cite{Cilindros}
\begin{equation}
\frac{d\vec{S}}{d\tau}=\frac{1}{2}\vec{S}\times\vec{H}\ ,\label{eq:SpinPrec}
\end{equation}
which is analogous to the precession of a magnetic dipole in a magnetic
field, $D\vec{S}/d\tau=\vec{\mu}\times\vec{B}$. When the electromagnetic
field is non-homogeneous, a force is also exerted on the magnetic
dipole, covariantly described by $DP^{\alpha}/d\tau=B^{\beta\alpha}\mu_{\beta}$
\cite{Costa:2012cy,Dixon1964}, where $\mu^{\beta}$ is the magnetic
dipole moment 4-vector, and $B_{\alpha\beta}=\star F_{\alpha\mu;\beta}U^{\mu}$
($F^{\alpha\beta}\equiv$ Faraday tensor, $\star\equiv$ Hodge dual)
is the ``magnetic tidal tensor'' as measured by the particle. A
covariant force is likewise exerted on a gyroscope in a gravitational
field (the ``spin-curvature'' force\cite{Mathisson:1937zz,Papapetrou:1951pa,Costa:2012cy,Dixon1964}),
which can be written in the remarkably similar form \cite{Costa:2012cy}
\begin{equation}
\frac{DP^{\alpha}}{d\tau}=-\mathbb{H}^{\beta\alpha}S_{\beta};\qquad\quad\mathbb{H}_{\alpha\beta}\equiv\star R_{\alpha\mu\beta\nu}U^{\mu}U^{\nu}=\frac{1}{2}\epsilon_{\alpha\mu}^{\ \ \ \lambda\tau}R_{\lambda\tau\beta\nu}U^{\mu}U^{\nu}.\label{eq:SpinCurvature}
\end{equation}
Here $\mathbb{H}_{\alpha\beta}$ is the ``gravitomagnetic tidal tensor''
(or ``magnetic part'' of the Riemann tensor) as measured by the
particle, playing a role analogous to that of $B_{\alpha\beta}$ in
electromagnetism. For a particle at rest in a stationary field in
the form (\ref{eq:StatMetric}), it is related to the gravitomagnetic
field $\vec{H}$ by the expression \cite{Analogies} 
\begin{eqnarray}
\mathbb{H}_{ij} & = & -\frac{1}{2}\left[\tilde{\nabla}_{j}H_{i}+(\vec{G}\cdot\vec{H})h_{ij}-2G_{j}H_{i}\right]\ .\label{eq:HijGEM}
\end{eqnarray}
In the linear regime, $\mathbb{H}_{ij}\approx H_{i,j}$, and so one
can say that (comparing to $\vec{H}$ ) $\mathbb{H}_{\alpha\beta}$
is essentially a quantity one order higher in differentiation of $\vec{\mathcal{A}}$.

\subsection{Sagnac effect}

By contrast with classical electromagnetism (where only the curl of
the magnetic vector potential, $\nabla\times\vec{A}=\vec{B}$, manifests
physically), and more like in quantum theory (where $\vec{A}$ manifests
itself in the so-called Aharonov-Bohm effect\cite{AharonovBohm}),
in General Relativity there are also gravitational effects governed
by the gravitomagnetic vector potential $\vec{\mathcal{A}}$ (or 1-form
$\bm{\mathcal{A}}$). One of them is the Sagnac effect, consisting
of the difference in arrival times of light-beams propagating around
a spatially closed path in opposite directions. In flat spacetime,
where the concept was first introduced (see e.g. Refs. \cite{Post1967,Kajari:2009qy}),
the time difference is originated by the rotation of the apparatus
with respect to global inertial frames (thus to the ``distant stars''),
see Fig. 1 in Ref. \cite{Cilindros}. In a gravitational field,
however, it arises also in apparatuses which are fixed relative to
the distant stars (i.e., to \emph{asymptotically} inertial frames),
Fig. \ref{fig:SagnacClockeffect}(a) below, signaling frame-dragging
\cite{AshtekarMagnon,Tartaglia:1998rh,Kajari:2009qy,MinguzziSimultaneity,Ruggiero_SagnacTestsReview,Cilindros}.
In both cases the effect can be read off from the spacetime metric
(\ref{eq:StatMetric}), encompassing the flat Minkowski metric written
in a rotating coordinate system, as well as arbitrary stationary gravitational
fields. Along a photon worldline, $ds^{2}=0$; by (\ref{eq:StatMetric}),
this yields two solutions, the future-oriented one being $dt=\mathcal{A}_{i}dx^{i}+e^{-\Phi}dl$,
where $dl=\sqrt{h_{ij}dx^{i}dx^{j}}$ is the spatial distance element.
Consider photons constrained to move within a closed loop $C$ in
the space manifold $\Sigma$; for instance, an optical fiber loop,
as depicted in Fig. \ref{fig:SagnacClockeffect} (a). Using the +
(-) sign to denote the anti-clockwise (clockwise) directions, the
coordinate time it takes for a full loop is, respectively, $t_{\pm}=\oint_{\pm C}dt=\oint_{C}e^{-\Phi}dl\pm\oint_{C}\mathcal{A}_{i}dx^{i}$;
therefore, the Sagnac \emph{coordinate} time delay $\Delta t_{{\rm S}}$
($\Delta t$ in the notation of Ref. \cite{Cilindros}) is 
\begin{equation}
\Delta t_{{\rm S}}\equiv t_{+}-t_{-}=2\oint_{C}\mathcal{A}_{i}dx^{i}=2\oint_{C}\bm{\mathcal{A}}\ ,\label{eq:SagnacDiffForm}
\end{equation}
translating, in the observer's proper time, to $\Delta\tau_{{\rm S}}=e^{\Phi}\Delta t_{{\rm S}}$.
We can thus cast gravitomagnetism into the three distinct levels in
Table \ref{tab:Levels}, corresponding to different orders of differentiation
of $\bm{\mathcal{A}}$ (the first one being $\bm{\mathcal{A}}$ itself).

\subsection{Synchronization gap\label{subsec:Synchronization-gap}}

Another physical process where $\bm{\mathcal{A}}$ manifests is in
the synchronization of the clocks carried by the ``laboratory observers''
(i.e., tangent to the Killing vector field $\partial_{t}$). Consider
a curve $x^{\alpha}(\lambda)$ of tangent $dx^{\alpha}/d\lambda$
which is spatially closed (i.e., its projection onto $\Sigma$ yields
a closed curve $C$, so that after each loop it re-intersects the
worldline of the original observer). Along $x^{\alpha}(\lambda)$,
the synchronization through Einstein's light signaling procedure \cite{LandauLifshitz,MinguzziSimultaneity}
amounts to the condition that the curve be orthogonal (at every point)
to $\partial_{t}^{\alpha}$, that is, $g_{0\beta}dx^{\beta}/d\lambda=0\Leftrightarrow dt=\mathcal{A}_{i}dx^{i}$.
This curve will thus re-intersect the worldline of the original observer
at a coordinate time $t_{f}=t_{i}+\Delta t_{{\rm sync}}$, where\cite{LandauLifshitz}
\begin{equation}
\Delta t_{{\rm sync}}=\oint_{C}\mathcal{A}_{i}dx^{i}=\oint_{C}\bm{\mathcal{A}}\quad\,(=\Delta t_{{\rm S}}/2)\ .\label{eq:syncgap}
\end{equation}
The observer will then find that his clock is not synchronized with
his preceding neighbor's by a time gap (as measured in his proper
time) $\Delta\tau_{{\rm sync}}=e^{\Phi}\Delta t_{{\rm sync}}$, corresponding,
in coordinate time, to an interval $\Delta t_{{\rm sync}}$ {[}which
is one half the Sagnac time delay $\Delta t_{{\rm S}}$ along such
loop, Eq. \eqref{eq:SagnacDiffForm}{]}. Only when $\Delta t_{{\rm sync}}=0$
the observers are able to fully synchronize their clocks along a closed
loop.

\subsection{Gravitomagnetic clock effect}

As is well known, in the field of a spinning body the periods of co-
and counter-rotating circular geodesics differ; such an effect has
been dubbed \cite{COHENMashhoon1993,BonnorClockEffect,BiniJantzenMashhoon_Clock1,IorioClockEffect}
gravitomagnetic ``clock effect''. The corresponding angular velocities
read (see e.g. Sec. 3.1 of Ref. \cite{Cilindros})
\begin{equation}
\Omega_{{\rm geo}\pm}=\frac{-g_{0\phi,r}\pm\sqrt{g_{0\phi,r}^{2}-g_{\phi\phi,r}g_{00,r}}}{g_{\phi\phi,r}}\ ,\label{eq:Omegageo}
\end{equation}
and thus the difference between their periods, $\Delta t_{{\rm geo}}=2\pi(\Omega_{{\rm geo+}}^{-1}+\Omega_{{\rm geo-}}^{-1})=-4\pi g_{0\phi,r}/g_{00,r}$,
is\cite{Cilindros,PaperDragging}
\begin{equation}
\Delta t_{{\rm geo}}=\Delta t_{{\rm S}}+\Delta t_{H};\qquad\Delta t_{{\rm S}}=4\pi\mathcal{A}_{\phi};\qquad\Delta t_{H}=-2\pi\frac{\star H_{r\phi}}{G_{r}e^{\Phi}}\ ,\label{eq:GMClock}
\end{equation}
where $\star H_{jk}\equiv\epsilon_{ijk}H^{i}$ is the 2-form dual
to the gravitomagnetic field $\vec{H}$, such that $\star H_{r\phi}=\sqrt{h}H^{z}$
in cylindrical coordinates, and $\star\!H_{r\phi}=-\sqrt{h}H^{\theta}$
in spherical coordinates. Here $h$ is the determinant of the space
metric $h_{ij}$ in \eqref{eq:StatMetric}. Hence, the effect consists
of the sum of two contributions of different origin, corresponding
to two distinct levels of gravitomagnetism in Table \eqref{tab:Levels}:
the Sagnac time delay $\Delta t_{{\rm S}}$ around the circular loop,
governed by $\bm{\mathcal{A}}$, cf. Eq. \eqref{eq:SagnacDiffForm},
plus the term $\Delta t_{H}$ due to the gravitomagnetic force $\gamma\vec{U}\times\vec{H}$
in Eq. \eqref{eq:QMGeo} (which has a direct electromagnetic counterpart,
see Sec. 3.1 in Ref. \cite{Cilindros}; cf. also Ref. \cite{IorioClockEffect}). 

The delay (\ref{eq:GMClock}) corresponds to orbital periods (in coordinate
time) as seen by the ``laboratory'' observers, at rest in the coordinates
of \eqref{eq:StatMetric}. Other observers (e.g. rotating with respect
to the former) will measure different periods since, from their point
of view, the closing of the orbits occurs at different points. An
\emph{observer-independent} akin effect\cite{Tartaglia_ClockI_CQG2000,BiniJantzenMashhoon_Clock1}
can however be derived, based on the proper times ($\tau_{+}$ and
$\tau_{-}$) measured by each orbiting particle between the events
where they meet, see Fig. \ref{fig:SagnacClockeffect} (b). Set a
starting meeting point at $\phi_{+}=\phi_{-}=0$, $t=0$; the next
meeting point is defined by $\phi_{+}=2\pi+\phi_{-}$. Since $\phi_{\pm}=\Omega_{{\rm geo}\pm}t$,
the meeting point occurs at a coordinate time $t=2\pi/(\Omega_{{\rm geo}+}-\Omega_{{\rm geo}-})$.
Hence, 
\begin{equation}
\tau_{\pm}=\frac{t}{U_{\pm}^{0}}=\frac{2\pi(U_{\pm}^{0})^{-1}}{\Omega_{{\rm geo}+}-\Omega_{{\rm geo}-}}\ ;\qquad\quad\Delta\tau\equiv\tau_{+}-\tau_{-}=2\pi\frac{(U_{+}^{0})^{-1}-(U_{-}^{0})^{-1}}{\Omega_{{\rm geo}+}-\Omega_{{\rm geo}-}}\ .\label{eq:InvClockEffect}
\end{equation}

\section{Artificial features of the usual form of the Weyl class metric\label{sec:Artificial-features-of}}

In the case of the static Levi-Civita cylinder, which follows from
\eqref{eq:LewisMetric}-\eqref{eq:LewisFunctions}, with $n$ and
$a$ real, by making\footnote{We shall see that the condition $c=0$ is actually not necessary,
cf. Eqs. \eqref{eq:Canonical}-\eqref{eq:Canonical2Functions}.} $b=0=c$, we have\footnote{Taking $a>0$, so that $t$ is the temporal coordinate.}
$\Phi=2\lambda_{{\rm m}}\ln(r)+const.$, ${\bf G}=-2\lambda_{{\rm m}}/r{\bf d}r$,
$\bm{\mathcal{A}}=0=\vec{H}$, where $\lambda_{{\rm m}}=(1-n)/4$.
This exactly matches the electromagnetic counterparts for an infinite
static charged cylinder, if ones identifies $\lambda_{{\rm m}}$ with
minus the charge per unit $z$-length. ($\lambda_{{\rm m}}$ being
actually the Komar mass per unit $z$- length, as we shall see in
Sec. \ref{subsec:CanonicalKomar}). However, in the general case $b\ne0\ne c$%
, we have
\begin{itemize}
\item $\Phi$, $\bm{\mathcal{A}}$, and $\vec{G}$ complicated {[}Eqs. (43)
of Ref. \cite{Cilindros}{]}, and very different from the electromagnetic
counterparts for a infinitely long spinning charged cylinder (in the
inertial rest frame);
\item $\vec{H}$ and $\mathbb{H}_{\alpha\beta}$ both \emph{non-zero} {[}and
complicated, Eqs. (43) and (45) of Ref. \cite{Cilindros}{]}, at
odds with the electromagnetic analogue.
\end{itemize}
These features are somewhat unexpected given the similarities with
the electromagnetic analogue in the static case, and given that this
metric is known to be \emph{locally static}. The situation resembles
more the electromagnetic analogue as seen from a rotating frame. Moreover,
\begin{itemize}
\item $\partial_{t}$ ceases to be time-like for $r^{2n}>a^{2}n^{2}/c^{2}$
$\Rightarrow$ no observers at rest are possible past this radius
\end{itemize}
which is, again, reminiscent of a rigidly rotating frame in flat spacetime
where, past a certain value of $r$, the observers would be superluminal.
The question then arises, can the metric, in the usual form \eqref{eq:LewisMetric}-\eqref{eq:LewisFunctions}
given in the literature, be actually written in some trivially rotating
coordinate system? We will next show this to be the case.

\section{The \textquotedblleft canonical\textquotedblright{} form of the Weyl
class metric}

\subsection{\textquotedblleft Star-fixed\textquotedblright{} coordinates: the
metric with only three parameters\label{subsec:Non-rotating-(star-fixed)-coordi}}

An analysis of the curvature invariants {[}cubic and quadratic, Eqs.
(39)-(41) and (50) of Ref. \cite{Cilindros}{]} reveals that the
Weyl class metric (i.e., with $n$ real) is a ``purely electric''
Petrov type I spacetime \cite{Cilindros,Invariants,McIntosh_1994}.
This means that, at each point, there is an (unique) observer for
which the gravitomagnetic tidal tensor vanishes, $\mathbb{H}_{\alpha\beta}=0$.
These observers have 4-velocity of the form $U^{\alpha}=U^{t}(\partial_{t}^{\alpha}+\Omega\partial_{\phi}^{\alpha})$,
with \emph{constant} angular velocity $\Omega$ given by
\begin{equation}
\Omega=\frac{c}{n-bc}\qquad{\rm or}\qquad\Omega=-\frac{1}{b}\ ,\label{eq:angvelocities}
\end{equation}
the first (second) value yielding a time-like $U^{\alpha}$ if $a>0$
($a<0$). Thus, by performing a coordinate rotation at constant angular
velocity
\begin{equation}
\phi=\varphi-\Omega t\ ,\label{eq:transformation}
\end{equation}
one switches to a coordinate system where these observers are at rest,
and the metric takes the form 
\begin{equation}
ds^{2}=-\frac{r^{1-s}}{\alpha}(dt+\bar{C}d\phi)^{2}+r^{(s^{2}-1)/2}(dr^{2}+dz^{2})+\alpha r^{1+s}d\phi^{2}\ ,\label{eq:Canonical}
\end{equation}
with, for $\Omega=c/(n-bc)$, 
\begin{equation}
s=n\ ;\qquad\alpha=\frac{\bar{C}^{2}}{ab^{2}}\ ;\qquad\bar{C}=b\frac{n-bc}{n}=b\frac{s-bc}{s}\ ,\label{eq:CanonicalFunctions}
\end{equation}
and, for $\Omega=-1/b$, 
\begin{equation}
s=-n\ ;\qquad\alpha=-ab^{2}\ ;\qquad\bar{C}=-b\frac{n-bc}{n}=b\frac{s-bc}{s}\ .\label{eq:Canonical2Functions}
\end{equation}
Equation \eqref{eq:Canonical} shows that the metric depends only
on three effective parameters: $\alpha$, $s$, and $\bar{C}$, manifesting
a redundancy in the original four parameters {[}different values of
$(n,a,b,c)$ yielding the same values of $(s,\alpha,\bar{C})$ correspond
to the same physical solution{]}. The two values \eqref{eq:angvelocities}
for the angular velocity $\Omega$ are two equivalent paths of reaching
\eqref{eq:Canonical}, and manifest a particular case of the redundancy:
two sets of parameters $(a_{1},b_{1},c_{1},n_{1})$ and $(a_{2},b_{2},c_{2},n_{2})$,
with $a_{1}>0$ and $a_{2}<0$, such that the values of $(s,\alpha,\bar{C})$
are the same. There is one special case excluded from each of the
transformations \eqref{eq:angvelocities}-\eqref{eq:transformation};
namely, $bc=n$ for the first value of $\Omega$, and $b=0$ for the
second; they are however redundant, as both lead to the Levi-Civita
line-element\footnote{That it is so for $b=0$ can be immediately seen by substituting in
(\ref{eq:Canonical})-(\ref{eq:CanonicalFunctions}), yielding \eqref{eq:LeviCivita};
likewise, that it is so for $n=bc$ can be seen by substituting $n\rightarrow bc$
in the expression for $\bar{C}$ in (\ref{eq:Canonical2Functions}).} \eqref{eq:LeviCivita}.

Observe that the Killing vector field $\partial_{t}$%
{} is, in (\ref{eq:Canonical}), \emph{everywhere time-like} ($g_{00}<0$
for all $r$). Therefore, observers of 4-velocity $u^{\alpha}=(-g_{00})^{1/2}\partial_{t}^{\alpha}$,
at rest in the coordinates of (\ref{eq:Canonical}), exist everywhere
(even for arbitrarily large $r$). As we shall see in Sec. \eqref{subsec:Physical-properties;-gravitomagn}
below, $\partial_{t}$ is actually tangent to inertial observers at
infinity, hence the reference frame associated to the coordinate system
in \eqref{eq:Canonical} is asymptotically inertial, and thus fixed
to the ``distant stars''.

\subsection{Symmetry under swap of time and angular coordinates}

In the transformation \eqref{eq:transformation} one assumes, as is
usual practice, that $\varphi$ is an angular coordinate, ranging
$[0,2\pi[$, and $t$ the time coordinate {[}which in turn implies
$\alpha>0$ and thus $a>0$ in case \eqref{eq:CanonicalFunctions},
and $a<0$ in case \eqref{eq:Canonical2Functions}{]}. Such assumption
is however not necessary to reach \eqref{eq:Canonical}. Indeed, swapping
$t\leftrightarrow\varphi$ in \eqref{eq:LewisMetric}, again leads
to \eqref{eq:Canonical}, as we shall now show. Substituting, in \eqref{eq:LewisMetric},
$\varphi\rightarrow t'$, $t\rightarrow\varphi'$, the time-like observers
measuring vanishing gravitomagnetic tidal tensor have now angular
velocity $\Omega'=1/\Omega$, where $\Omega$ is given by \eqref{eq:angvelocities},
and $\Omega'=(n-bc)/c$ yields a time-like $U^{\alpha}$ if $a>0$,
and likewise $\Omega'=-b$ for $a<0$. Applying the transformation
$\phi'=\varphi'-\Omega't'=\varphi'-t'/\Omega$ to such line element
leads to a primed version of \eqref{eq:Canonical}, 
\begin{equation}
ds^{2}=-\frac{r^{1-s}}{\alpha}(dt'+\bar{C}d\phi')^{2}+r^{(s^{2}-1)/2}(dr^{2}+dz^{2})+\alpha r^{1+s}d\phi'^{2}\ ,\label{eq:SwappedOriginal}
\end{equation}
with, for $\Omega'=(n-bc)/c$, the identifications $s=n$, $\alpha=c^{2}/(an^{2})$,
$\bar{C}=c/n$; and, for $\Omega'=-b$, the identifications $s=-n$,
$\alpha=-a$, $\bar{C}=-c/n$.

One must note, however, that the metric in star fixed coordinates
\eqref{eq:Canonical} does not preserve such symmetry. The coordinate
rotation \eqref{eq:transformation}, with the identification $\varphi=\varphi+2\pi$,
breaks that symmetry by implicitly choosing $\phi$ (and $\varphi$)
as a periodic coordinate, and $t$ non-periodic. Indeed, substituting
in \eqref{eq:Canonical} $\phi\rightarrow t'$, $t\rightarrow\phi'$,
leads to \eqref{eq:SwappedOriginal} with $\phi'\leftrightarrow t'$
swapped: 
\begin{equation}
ds^{2}=-\frac{r^{1-s}}{\alpha}(\bar{C}dt'+d\phi')^{2}+r^{(s^{2}-1)/2}(dr^{2}+dz^{2})+\alpha r^{1+s}dt'^{2}\ ;\label{eq:SwappForbidden}
\end{equation}
forcing now on it the identification $(t',\phi')=(t',\phi'+2\pi)$
(i.e, taking $\phi'$ to be periodic), makes it become the Levi-Civita
metric in a rotating coordinate system, immediately diagonalizable
through the coordinate rotation $\phi''=\phi'+\bar{C}t$. This occurs
because, by overriding the original identifications $(t,\phi)=(t,\phi+2\pi)$,
the geometry was \emph{globally} (albeit not locally) changed. Indeed
such transformation (with such identifications) is not a global diffeomorphism,
as can be seen e.g. from the fact the ordered pairs $\mathcal{P}_{1}$:
$(t,\phi)$ and $\mathcal{P}_{2}$: $(t,\phi+2\pi)$, which represented
the same event in the original metric \eqref{eq:Canonical}, are mapped
into the two different events $\mathcal{P}'_{1}$: $(t',\phi')$ and
$\mathcal{P}'_{2}$: $(t'+2\pi,\phi')$ in the metric \eqref{eq:SwappForbidden}.
The transformation $(t',\phi')=(\phi,t)$, followed by $\phi''=\phi'+\bar{C}t$,
actually amounts to (76) of Ref. \cite{Cilindros} which, as shown
therein, corresponds to the ``famous'' \cite{SantosGRG1995,GriffithsPodolsky2009}
transformation that takes the Weyl class metric into the static Levi-Civita
one.

\subsection{The metric in terms of physical parameters --- \textquotedblleft canonical\textquotedblright{}
form\label{subsec:CanonicalKomar}}

The fact that in Eq. \eqref{eq:Canonical} the Killing vector field
$\xi^{\alpha}=\partial_{t}^{\alpha}$ is everywhere time-like, tangent
to inertial observers at infinity, and appropriately normalized \cite{Cilindros},
allows for defining a corresponding Komar integral on simply connected
tubes $\mathcal{V}$ of unit $z$-length parallel to the $z$-axis,
having a physical interpretation of ``active'' gravitational mass
per unit $z$-length, as discussed in Secs. 5.2.1 and 2.4 of Ref.
\cite{Cilindros} (cf. also Refs. \cite{Bonnor1979_linemass,Israel1977,Marder1958,BicakKomar}).
It is given by
\[
\lambda_{{\rm m}}=\frac{1}{8\pi}\int_{\partial\mathcal{V}}\star\mathbf{d}\bm{\xi}=\frac{1}{8\pi}\int_{\mathcal{S}}(\star d\xi)_{\phi z}d\phi dz=\frac{1-s}{4}\ ,
\]
where $\partial\mathcal{V}=\mathcal{S}\cup\mathcal{B}_{1}\cup\mathcal{B}_{2}$
is the tube's boundary, $\mathcal{S}$ its lateral surface, parameterized
by $\{\phi,z\}$, and $\mathcal{B}_{1}$ and $\mathcal{B}_{2}$ its
bases, lying on the planes orthogonal to the $z$-axis and parameterized
by $\{r,\phi\}$, and in the second and third equalities we noticed
that $(\star d\xi)_{r\phi}=0$ and $(\star d\xi)_{\phi z}=1-s$. Likewise,
the Komar integral associated with the axial symmetry Killing vector
field $\zeta^{\alpha}=\partial_{\phi}^{\alpha}$ has the interpretation
of the spacetime's angular momentum per unit $z$-length,
\[
j=-\frac{1}{16\pi}\int_{\partial\mathcal{V}}\star\mathbf{d}\bm{\zeta}=-\frac{1}{16\pi}\int_{\mathcal{S}}(\star d\zeta)_{\phi z}d\phi dz=\frac{s\bar{C}}{4}\ ,
\]
where in the second equality we noticed that $(\star\zeta)_{r\phi}=0$.
It follows that \eqref{eq:Canonical} can be re-written as

\begin{equation}
\boxed{ds^{2}=-\frac{r^{4\lambda_{{\rm m}}}}{\alpha}\left(dt-\frac{j}{\lambda_{{\rm m}}-1/4}d\phi\right)^{2}+r^{4\lambda_{{\rm m}}(2\lambda_{{\rm m}}-1)}(dr^{2}+dz^{2})+\alpha r^{2(1-2\lambda_{{\rm m}})}d\phi^{2}\ .}\label{eq:MetricKomar}
\end{equation}

\subsection{Physical properties; gravitomagnetism\label{subsec:Physical-properties;-gravitomagn}}

For $\alpha>0$ {[}so that $t$ in Eq. (\ref{eq:MetricKomar}) is
a temporal coordinate{]}, the metric can be put in the form (\ref{eq:StatMetric}),
with
\begin{align}
 & e^{2\Phi}=\frac{r^{4\lambda_{{\rm m}}}}{\alpha}\quad\Rightarrow\quad\Phi=2\lambda_{{\rm m}}\ln(r)+K\ ;\qquad\quad\bm{\mathcal{A}}=-\frac{4j}{1-4\lambda_{{\rm m}}}{\bf d}\phi\ ;\label{eq:CanonicalQM1}\\
 & h_{ij}dx^{i}dx^{j}=r^{4\lambda_{{\rm m}}(2\lambda_{{\rm m}}-1)}(dr^{2}+dz^{2})+\alpha r^{2(1-2\lambda_{{\rm m}})}d\phi^{2}\label{eq:CanonicalQM2}
\end{align}
{[}with $K\equiv-\ln(\alpha)/2${]}. The gravitoelectric and gravitomagnetic
1-forms/fields read, cf. Eqs. (\ref{eq:QMGeo}), 
\begin{equation}
{\bf G}=-\frac{2\lambda_{{\rm m}}}{r}{\bf d}r\ ;\qquad\vec{H}={\bf H}=0\ .\label{eq:GEMCanonical}
\end{equation}
Thus $\Phi$, ${\bf G}$, and ${\bf H}=0$ match their electromagnetic
counterparts for a spinning charged cylinder (as viewed from the inertial
rest frame, see e.g. Sec. 4 in Ref. \cite{Cilindros}) identifying
the Komar mass with \emph{minus} the charge, $\lambda_{{\rm m}}\leftrightarrow-\lambda$;
the gravitomagnetic potential 1-form $\bm{\mathcal{A}}=\mathcal{A}_{\phi}{\bf d}\phi$
also resembles the magnetic potential 1-form $\mathbf{A}=\mathfrak{m}\mathbf{d}\phi$.
The cylinder's rotation does not manifest in the inertial forces (nor
in the tidal forces, as shown in Sec. 5.2.3 of Ref. \cite{Cilindros});
the only inertial force acting on test particles is the gravitoelectric
(Newtonian-like) force $m\vec{G}$, independent of $j$. Therefore,
particles dropped from rest, or in initial radial motion, move along
radial straight lines, cf. Eq. (\ref{eq:QMGeo}); and circular geodesics
have a \emph{constant} speed given by $v_{{\rm geo}}=\sqrt{\lambda_{{\rm m}}/(1/2-\lambda_{{\rm m}})}$,
being thus possible when $0\le\lambda_{{\rm m}}<1/4$ (it is when
$\lambda_{{\rm m}}>0$ that $\vec{G}$ is attractive, and they become
null for $\lambda_{{\rm m}}=1/4$).\textcolor{blue}{{} }The vanishing
of $\vec{H}$ means also that gyroscopes at rest in the coordinates
of (\ref{eq:MetricKomar}) do not precess, the components of their
spin vector $\vec{S}$ remaining constant, cf. Eq. (\ref{eq:SpinPrec}).
Since $\vec{G}\stackrel{r\rightarrow\infty}{\rightarrow}\vec{0}$,
it follows moreover that the reference frame associated to the coordinate
system in (\ref{eq:MetricKomar}) is \emph{asymptotically} inertial,
thus one can take it as the rest frame of the ``distant stars''
(``star-fixed'' frame).

\begin{figure}
\includegraphics[width=1\textwidth]{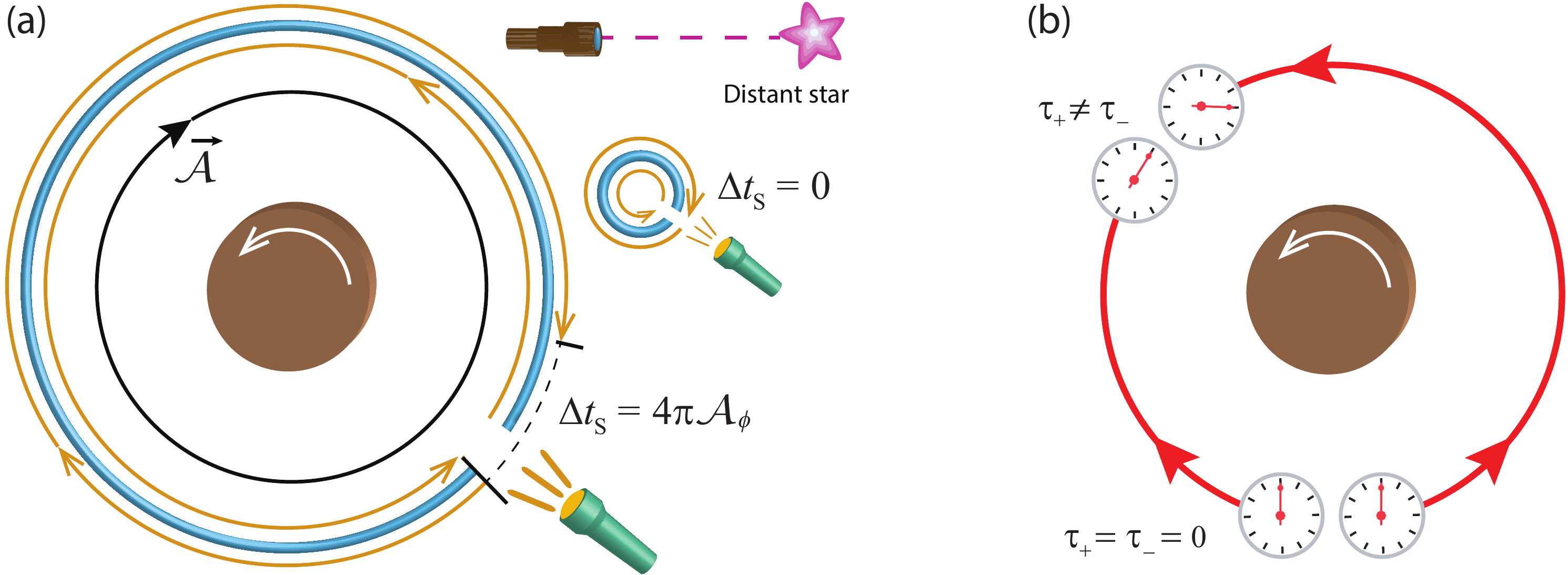}\caption{\label{fig:SagnacClockeffect}(a) Sagnac effect around spinning cylinders:
a flashlight sends light beams propagating in opposite directions
along optical fiber loops fixed with respect to the \textquotedblleft distant
stars\textquotedblright{} (i.e., to the asymptotic inertial frame at
infinity). In each loop $C$, the difference in beam arrival times
is $\Delta t_{{\rm S}}=2\oint_{C}\bm{\mathcal{A}}$; since $\bm{\mathcal{A}}$
is a \emph{closed} form (${\bf d}\bm{\mathcal{A}}=0$), the effect
vanishes for all loops not enclosing the cylinder, and has the same
value \eqref{eq:DtCanonical} for all loops enclosing it (the co-rotating
beam arriving first). (b) Frame independent gravitomagnetic clock
effect: a pair of clocks in oppositely rotating circular geodesics
around a cylinder; when the cylinder spins ($j\protect\ne0$) the
clocks measure different proper times between the events where they
meet, $\tau_{+}>\tau_{-}$.}
\end{figure}

The only surviving gravitomagnetic object is thus the gravitomagnetic
1-form $\bm{\mathcal{A}}$, corresponding to the first level in Table
\ref{tab:Levels}. One of its physical manifestations is the Sagnac
effect: consider, as depicted in Fig. \ref{fig:SagnacClockeffect}(a)
optical fiber loops fixed with respect to the distant stars, i.e.,
at rest in the coordinate system of (\ref{eq:MetricKomar}). The difference
in arrival times for light beams propagating in opposite directions
along any of such loops is given by the circulation of $\bm{\mathcal{A}}$
along the loop, c.f. Eq. \eqref{eq:SagnacDiffForm}. Observe that
$\bm{\mathcal{A}}$ is a closed form, $\mathbf{d}\bm{\mathcal{A}}=0$
(since $\mathcal{A}_{\phi}$ is constant), defined in a space manifold
$\Sigma$ homeomorphic to $\mathbb{R}^{3}\backslash\{r=0\}$. By the
Stokes theorem, this means\cite{Cilindros} that the effect vanishes
along any loop which does not enclose the central cylinder (or the
axis $r=0$), such as the small loop in Fig. \ref{fig:SagnacClockeffect}
(a), and has the \emph{same} nonzero value 
\begin{equation}
\Delta t_{{\rm S}}=4\pi\mathcal{A}_{\phi}=-\frac{4\pi j}{1/4-\lambda_{{\rm m}}}\label{eq:DtCanonical}
\end{equation}
along any loop enclosing the cylinder, regardless of its shape; for
instance, the large circular loop depicted in Fig.~\ref{fig:SagnacClockeffect}(a).
It is worth noticing that this mirrors the situation for the Aharonov-Bohm
effect around spinning charged cylinders, which is likewise independent
of the shape of the paths; the two effects are actually described
by formally analogous equations\footnote{Re-writing \eqref{eq:SagnacDiffForm} in terms of the half-loop phase
delay $\Delta\varphi=(E/\hbar)\Delta t_{{\rm S}}/2=(2\pi E/\hbar)\mathcal{A}_{\phi}$
and identifying $\{E,\mathcal{A}_{\phi}\}\leftrightarrow\{q,A_{\phi}\}$,
where $E\equiv$ photon's energy, see Sec. 4.1 in Ref. \cite{Cilindros}.}. 

The apparatus above makes use of a star-fixed reference frame, which
is physically realized by aiming telescopes at the distant stars\cite{CiufoliniWheeler,CiufoliniNature2007}.
It is possible, however (still based on the Sagnac effect), to detect
the cylinder's rotation in a more local way, without the need for
setting up a specific frame; only not with a single loop, as along
a single loop the effect can always be made to vanish by spinning
it with some angular velocity. In particular, for a concentric circular
loop, the effect vanishes if it has zero angular momentum, i.e., if
it comoves with the zero angular momentum observers (ZAMOs) of the
same radius. The angular velocity of such observers, Eq. (69) of Ref.
\cite{Cilindros}, is however $r$-dependent; hence, considering
instead a ``coil'' of optical loops%
, as depicted in Fig. 4 of Ref. \cite{Cilindros}, provides a frame-independent
(thought) experiment to detect the cylinder's rotation, since it is
impossible to make the effect vanish simultaneously in every loop
when $j\ne0\Leftrightarrow\text{\ensuremath{\bm{\mathcal{A}}}}\ne0$.

\subsubsection{Observer-independent gravitomagnetic clock effect}

Another consequence of the vanishing of $\vec{H}$ is that the gravitomagnetic
clock effect in Eq. (\ref{eq:GMClock}) reduces to the Sagnac time
delay, $\Delta t_{{\rm geo}}=\Delta t_{{\rm S}}=4\pi\mathcal{A}_{\phi}$;
hence, all that was said above about beams in optical loops around
the cylinder, applies as well to pairs of particles in oppositely
rotating circular geodesics (the co-rotating geodesics having thus
shorter periods). It is however actually possible to detect the cylinder's
rotation using only \emph{one pair} of particles (i.e., a pair of
clocks), through the difference in the proper times ($\tau_{+}$ and
$\tau_{-}$) measured by each of them between the events where they
meet, see Fig. \ref{fig:SagnacClockeffect} (b). From Eqs. \eqref{eq:InvClockEffect}
and \eqref{eq:Omegageo}, with $U_{\pm}^{0}=(-g_{00}-2\Omega_{{\rm geo}\pm}g_{0\phi}-\Omega_{{\rm geo}\pm}^{2}g_{\phi\phi})^{-1/2}$,
we have
\[
\Delta\tau\equiv\tau_{+}-\tau_{-}=\frac{8\pi jr^{2\lambda_{{\rm m}}}}{\sqrt{\alpha(1-4\lambda_{{\rm m}})(1-2\lambda_{{\rm m}})+8\lambda_{{\rm m}}j^{2}r^{8\lambda_{{\rm m}}-2}\alpha^{-1}(1/4-\lambda_{{\rm m}})^{-1}}}\ \ (>0)
\]
(this result is mentioned in main paper \cite{Cilindros}, though
without presenting it explicitly). Hence, when $j\ne0$, the proper
times measured by each clock differ when they meet, the co-rotating
clock measuring a longer time. 

\subsection{Important limits: Levi-Civita static cylinder and cosmic strings}

It is immediate to obtain important limits from the canonical form
\eqref{eq:MetricKomar}. Taking the limit $j\rightarrow0$ yields
the Levi-Civita metric\cite{GriffithsPodolsky2009,SantosGRG1995,SantosCQG1995}
\begin{equation}
{\displaystyle {\displaystyle ds^{2}}=-\frac{r^{4\lambda_{{\rm m}}}}{\alpha}dt^{2}+r^{4\lambda_{{\rm m}}(2\lambda_{{\rm m}}-1)}(dr^{2}+dz^{2})}+\alpha r^{2(1-2\lambda_{{\rm m}})}d\phi^{2}\ .\label{eq:LeviCivita}
\end{equation}
The inertial fields $\vec{G}$ and $\vec{H}=0$, as well as the spatial
metric $h_{ij}$, remain the same as in \eqref{eq:MetricKomar} (the
same applying to the tidal fields/forces, see Sec. 5.2.3 in Ref. \cite{Cilindros}).
They differ only in the gravitomagnetic potential 1-form $\bm{\mathcal{A}}=-4j/(1-4\lambda_{{\rm m}})\mathbf{d}\phi$,
governing global physical effects such as the Sagnac effect and synchronization
gap \eqref{eq:syncgap} in loops around the cylinder, and the gravitomagnetic
clock effect, which are all zero for the static metric \eqref{eq:LeviCivita},
see Figs. \ref{fig:SagnacClockeffect}-\ref{fig:Hypersurfaces}.

The limit $\lambda_{{\rm m}}\rightarrow0$ yields 
\begin{equation}
{\displaystyle {\displaystyle ds^{2}}=-\frac{1}{\alpha}\left[dt+4jd\phi\right]^{2}+dr^{2}+dz^{2}}+\alpha r^{2}d\phi^{2}\label{eq:String}
\end{equation}
which is the metric of a spinning cosmic string\cite{SantosGRG1995,Barros_Bezerra_Romero2003,JensenSoleng}
of \emph{Komar} angular momentum per unit length $j$ and angle deficit
$2\pi(1-\alpha^{1/2})\equiv2\pi\delta$. In this case the spacetime
is locally flat ($R_{\alpha\beta\gamma\delta}=0$) for $r\ne0$. All
the GEM inertial fields vanish, $\vec{G}=\vec{H}=0$ (and the same
for the tidal fields), thus there are no gravitational forces of any
kind. Global gravitational effects however subsist, governed by $\bm{\mathcal{A}}=-4j\mathbf{d}\phi$
and $\alpha$. The non-vanishing gravitomagnetic potential 1-form
$\bm{\mathcal{A}}$ means that a Sagnac effect remains, thus the apparatuses
manifesting the source's rotation discussed in Sec.~\ref{subsec:Physical-properties;-gravitomagn}
apply here as well. The same applies to the synchronization of clocks:
observers at rest in the coordinates of \eqref{eq:String} (which
are in this case inertial observers) can synchronize their clocks
along closed loops not enclosing the string (i.e., the axis $r=0$),
but are unable to do so for loops enclosing it. As for the gravitomagnetic
clock effect, it does not apply here, as circular geodesics do not
exist (since there is no gravitational attraction, $\vec{G}=0$).
The angle deficit generates double images of objects located behind
the string \cite{KibbleCosmicStrings,Ford_Vilenkin_1981}, and a holonomy\cite{Nouri-Zonoz:2013rfa,Ford_Vilenkin_1981}
along closed (in spacetime or only spatially) loops around the string.
Namely, vectors parallel transported along such loops turn out rotated
by an angle $-2\pi\alpha^{1/2}$ (i.e, $2\pi\delta$) about the $z$-axis
when they return to the initial position --- an effect which is independent
of the shape of the loop and of $j$; see Sec. 5.2.4 of Ref. \cite{Cilindros}.
One thus can say that the metric \eqref{eq:String} possesses two
holonomies: a spatial holonomy governed by $\alpha$, which is the
same for spinning or non-spinning strings, plus a synchronization
holonomy (Sec. 5.3.3 of Ref. \cite{Cilindros}) that arises in
the spinning case. 

\subsection{Summary of \textquotedblleft canonical\textquotedblright{} features}

We argue Eq. (\ref{eq:MetricKomar}) to be the most natural, or \emph{canonical},
form for the metric of a Weyl class rotating cylinder for the following
reasons: 
\begin{itemize}
\item the Killing vector field $\partial_{t}$ is (for $\alpha>0$) everywhere
time-like (i.e., $g_{00}<0$ for all $r$), therefore physical observers
$u^{\alpha}=(-g_{00})^{-1/2}\partial_{t}^{\alpha}$, at rest in the
coordinates of (\ref{eq:MetricKomar}), exist everywhere. 
\item The associated reference frame is \emph{asymptotically} inertial,
and thus fixed with respect to the ``distant stars'' (Sec. \ref{subsec:Physical-properties;-gravitomagn}). 
\item A conserved Komar mass per unit length ($\lambda_{{\rm m}}$) can
be defined from $\partial_{t}$ which matches its expected value from
the gravitational field $\vec{G}$ and potential $\Phi$ in Sec. \ref{subsec:Physical-properties;-gravitomagn}
(see also Sec. 5.2.1 of Ref. \cite{Cilindros}), and also that
of the Levi-Civita static cylinder \eqref{eq:LeviCivita}. 
\item It is irreducibly given in terms of three parameters with a clear
physical interpretation: the Komar mass ($\lambda_{{\rm m}}$) and
angular momentum ($j$) per unit length, plus the parameter $\alpha$
governing the angle deficit of the spatial metric $h_{ij}$. 
\item The GEM fields are strikingly similar to the electromagnetic analogues
--- the electromagnetic fields of a rotating cylinder as measured
in the inertial rest frame (namely $\bm{\mathcal{A}}=\mathcal{A}_{\phi}\mathbf{d}\phi$;
$\mathcal{A}_{\phi}\equiv$constant, $\vec{H}=\mathbb{H}_{\alpha\beta}=0$,
and $\Phi$ and $G_{,i}$ match the electromagnetic counterparts identifying
the Komar mass per unit length $\lambda_{{\rm m}}$ with the minus
charge per unit length $\lambda$, cf. Sec. 4 of Ref. \cite{Cilindros}). 
\item It is immediate from it to obtain the two important limits: spinning
cosmic string ($\lambda_{{\rm m}}=0$), and Levi-Civita static solution
(evincing that $j=0$ is the necessary and \emph{sufficient} condition).
\item The GEM inertial fields and tidal tensors are the \emph{same }as those
of the Levi-Civita static cylinder (just like the electromagnetic
forces produced by a charged spinning cylinder are the same as by
a static one). 
\item It is obtained from a simple rigid rotation of coordinates, Eq. (\ref{eq:transformation}),
which is a well-defined \emph{global} coordinate transformation (Sec.
\ref{subsec:Non-rotating-(star-fixed)-coordi}). 
\item It makes immediately transparent the locally static but globally stationary
nature of the metric (see Sec. \ref{sec:Contrast-with-Kerr} below). 
\item It has a smooth matching to the van Stockum interior solution (corresponding
to a cylinder of rigidly rotating dust) \emph{written} \emph{in star-fixed
coordinates} (Sec. 5.4.2 of Ref. \cite{Cilindros}).
\end{itemize}
We conclude that the Lewis metric in its usual form (\ref{eq:LewisMetric})-(\ref{eq:LewisFunctions})
indeed possesses a trivial coordinate rotation {[}of angular velocity
$-\Omega$, equivalently given by either of Eqs. (\ref{eq:angvelocities}){]},
which has apparently gone unnoticed in the literature, and causes
the artificial features listed in Sec. \ref{sec:Artificial-features-of}.
As shown in Sec. 5.4 of Ref. \cite{Cilindros}, such rotation has
a simple interpretation when the solution is matched to the van Stockum
interior solution (corresponding to a rigidly rotating cylinder of
dust): the coordinate system in (\ref{eq:LewisMetric})-(\ref{eq:LewisFunctions})
is \emph{rigidly comoving} with the cylinder.

\section{Contrast with a locally (and globally) \emph{non-static} solution
--- the Kerr spacetime \label{sec:Contrast-with-Kerr}}

Question by O. Semer\'{a}k: \emph{you were comparing the results
for the (rotating) Weyl class Lewis metric with the static case; how
about the comparison with Kerr, which is different because there the
vorticity should contribute to the gravitomagnetic field?}

The contrast with the Kerr spacetime is indeed instructive. In what
pertains to gravitomagnetism, it fundamentally differs from the Weyl
class cylindrical metrics (rotating or non-rotating) in two mains
aspects: it is not locally static, and its Riemann tensor is not (except
at the equatorial plane) ``purely electric'' \cite{Invariants}.

\emph{Staticity.}--- a spacetime is static\cite{Stachel:1981fg}
within some region \emph{iff} a time-like Killing vector field $\xi^{\alpha}$
exists which is proportional to the gradient of some (single-valued)
function $\psi$, $\xi_{\alpha}=\eta\partial_{\alpha}\psi$. Locally,
this condition is equivalent to the integral lines of $\xi^{\alpha}$
having no vorticity, i.e., being hypersurface orthogonal (globally,
however, the vorticity-free condition is not sufficient \cite{Stachel:1981fg,Bonnor1980}).
One can show (Proposition 5.1 in Ref. \cite{Cilindros}) that,
in the GEM framework, local staticity amounts to the existence of
a coordinate system where the metric takes the stationary form \eqref{eq:StatMetric}
with $\bm{\mathcal{A}}$ closed ($\mathbf{d}\bm{\mathcal{A}}=0$);
and global staticity to $\bm{\mathcal{A}}$ being moreover an \emph{exact}
form (in a globally well defined coordinate system). 

\begin{figure}
\includegraphics[width=1\textwidth]{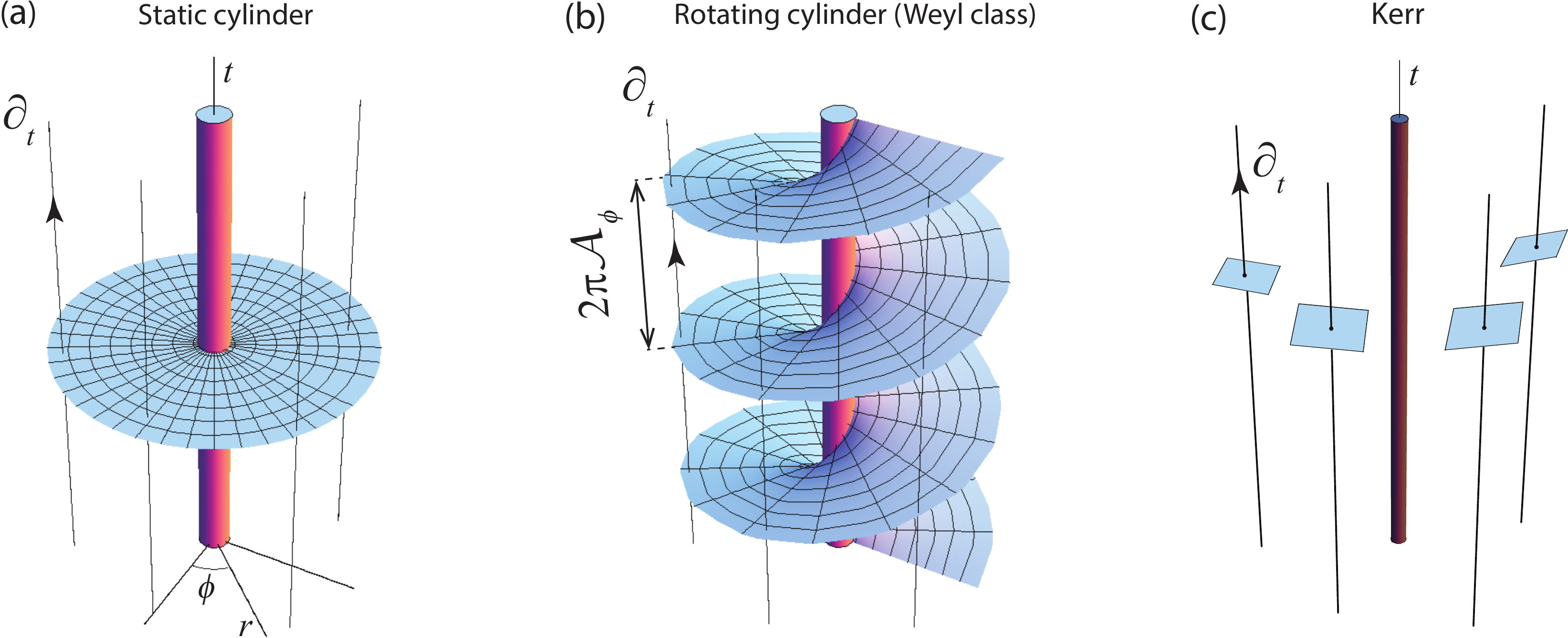}

\caption{\label{fig:Hypersurfaces}(a)-(b): In the Weyl class Lewis metrics
\eqref{eq:MetricKomar} the Killing vector field $\partial_{t}$ is
hypersurface orthogonal; such hypersurface is of global simultaneity
(a plane, in a $t,r,\phi$ plot) for a non-spinning (Levi-Civita)
cylinder, and of local but \emph{non-global} simultaneity (the helicoid
$t-\mathcal{A}_{\phi}\phi=const.$) in the spinning case. (c) In the
Kerr spacetime $\partial_{t}$ is not hypersurface orthogonal, i.e.,
the distribution of hyperplanes orthogonal to $\partial_{t}$ (hyperplanes
of \emph{local} simultaneity) is not integrable. In (a) observers
of worldlines tangent to $\partial_{t}$ (\textquotedblleft laboratory
observers\textquotedblright ) are able to globally synchronize their
clocks. In (b) they are unable to synchronize their clocks \emph{around}
the cylinder: each $2\pi$ turn along the helicoid leads to a different
event in time, the jump between turns being the synchronization gap
$\Delta t_{{\rm sync}}=2\pi\mathcal{A}_{\phi}$. In (c) the laboratory
observers are (generically) unable to synchronize their clocks along
any spatially closed loop, $\Delta t_{{\rm sync}}=\oint\bm{\mathcal{A}}\protect\ne0$.}
\end{figure}

The Weyl class Lewis metric \eqref{eq:MetricKomar} is locally static
since $\mathbf{d}\bm{\mathcal{A}}=0$; but, unless $j=0\Rightarrow\bm{\mathcal{A}}=0$
(Levi-Civita static cylinder), \emph{not globally} static, since $\bm{\mathcal{A}}=\mathcal{A}_{\phi}\mathbf{d}\phi$
is not an exact form. This means that the Killing vector field $\partial_{t}$
is hypersurface orthogonal but (unless $j=0$) such hypersurfaces
are not of global simultaneity, see Fig. \ref{fig:Hypersurfaces}
(a)-(b). In the case of the Kerr spacetime, $\mathbf{d}\bm{\mathcal{A}}\ne0$,
so it is not globally static; no hypersurface orthogonal time-like
Killing vector field exists, the only Killing vector field which is
time-like at infinity being $\partial_{t}$ in Boyer-Lindquist coordinates,
whose integral lines are well known to have vorticity. Geometrically,
this means that the distribution of hyperplanes orthogonal to $\partial_{t}$
(i.e., the hyperplanes of \emph{local} simultaneity\cite{MinguzziSimultaneity},
or local rest spaces of the ``laboratory'' observers) is not integrable,
see Fig. \ref{fig:Hypersurfaces} (c). On top of this, outside the
equatorial plane, $R_{\alpha\beta\gamma\delta}$ is not purely electric
(see Sec. V.C of Ref. \cite{Invariants}), thus no observers exist
measuring a vanishing gravitomagnetic tidal tensor $\mathbb{H}_{\alpha\beta}$.

Physically, this means that whereas for the Weyl class spinning cylinder
\eqref{eq:MetricKomar} only the first level of gravitomagnetism in
Table \ref{tab:Levels} is non-zero, in the Kerr spacetime all the
three levels are non-zero. Therein it is thus possible to detect the
source's rotation in a more local way (i.e., not needing experiments
on loops around the source): in a reference frame fixed to the distant
stars, due to the non-zero gravitomagnetic field $\vec{H}$, test
particles in geodesic motion will appear to be deflected by a gravitomagnetic
(or Coriolis) force $\gamma\vec{U}\times\vec{H}$, cf. Eq. \eqref{eq:QMGeo},
causing e.g. their orbits to precess (Lense-Thirring precession\cite{CiufoliniWheeler}),
and gyroscopes will as well be seen to precess, cf. Eq. \eqref{eq:SpinPrec}.
The non-vanishing $\mathbb{H}_{\alpha\beta}$ means moreover that
gyroscopes at rest (or generically moving) will be acted by a force
\eqref{eq:SpinCurvature}. 
\begin{table*}
\begin{tabular}{|c|l|c|c|}
\hline 
\multicolumn{2}{|c|}{\raisebox{2.5ex}{}\raisebox{0.2ex}{\textbf{\footnotesize{}Level
of Gravitomagnetism}}} & \multirow{2}{*}{{\footnotesize{}Lewis-Weyl}} & \multirow{2}{*}{{\footnotesize{}Kerr}}\tabularnewline
\cline{1-2} \cline{2-2} 
\raisebox{2.4ex}{}\raisebox{0.2ex}{{\footnotesize{}Governing object}} & \raisebox{0.2ex}{{\footnotesize{}Physical effect}} &  & \tabularnewline
\hline 
\raisebox{6.5ex}{}%
\begin{tabular}{c}
{\small{}$\vec{\mathcal{A}}$}\tabularnewline
{\footnotesize{}{}(gravitomagnetic}\tabularnewline
\raisebox{0.5ex}{{\footnotesize{}{}vector potential)}}\tabularnewline
\end{tabular} & \raisebox{1ex}{%
\begin{tabular}{l}
{\footnotesize{}$\bullet$ Sagnac effect}\tabularnewline
{\footnotesize{}$\bullet$ Synchronization gap}\tabularnewline
\end{tabular}} & %
\begin{tabular}{c}
{\Large{}$\checkmark$}\tabularnewline
{\footnotesize{}(global effects)}\tabularnewline
\end{tabular} & %
\begin{tabular}{c}
{\Large{}$\checkmark$}\tabularnewline
{\footnotesize{}(global }\emph{\footnotesize{}and}{\footnotesize{}
local)}\tabularnewline
\end{tabular}\tabularnewline
\hline 
\raisebox{3.5ex}{%
\begin{tabular}{c}
\raisebox{0.1ex}{{\small{}$\vec{H}$}}\tabularnewline
{\footnotesize{}{}(gravitomagnetic}\tabularnewline
{\footnotesize{}{}field $=e^{\phi}\nabla\times\vec{\mathcal{A}}$)}\tabularnewline
\end{tabular}} & \raisebox{2.5ex}{%
\begin{tabular}{l}
$\bullet$ {\footnotesize{}gravitomagnetic force}\tabularnewline
\raisebox{0.5ex}{{\footnotesize{}~~~$m\gamma\vec{U}\times\vec{H}$}}\tabularnewline
$\bullet$ {\footnotesize{}gyroscope precession}\tabularnewline
\raisebox{0.5ex}{{\footnotesize{}~~~$d\vec{S}/d\tau=\vec{S}\times\vec{H}/2$}}\tabularnewline
$\bullet$ {\footnotesize{}local Sagnac effect in}\tabularnewline
\raisebox{0.5ex}{~~~{\footnotesize{}light gyroscope}}\tabularnewline
\end{tabular}} & %
\begin{tabular}{c}
{\Large{}$\times$}\tabularnewline
\tabularnewline
\end{tabular} & %
\begin{tabular}{c}
{\Large{}$\checkmark$}\tabularnewline
\tabularnewline
\end{tabular}\tabularnewline
\hline 
{\small{}$\vec{H}+\vec{\mathcal{A}}$} & %
\begin{tabular}{l}
{\footnotesize{}$\bullet$ Gravitomagnetic}\tabularnewline
{\footnotesize{}~~~~``clock'' effect}\tabularnewline
\end{tabular} & %
\begin{tabular}{l}
{\footnotesize{}$\bullet$ co-rotating geodesic}\tabularnewline
{\footnotesize{}~~~has }\emph{\footnotesize{}shorter}{\footnotesize{}
period}\tabularnewline
\end{tabular} & %
\begin{tabular}{l}
{\footnotesize{}$\bullet$ co-rotating geodesic}\tabularnewline
{\footnotesize{}~~~has }\emph{\footnotesize{}longer}{\footnotesize{}
period}\tabularnewline
\end{tabular}\tabularnewline
\hline 
\raisebox{7ex}{}\raisebox{1ex}{%
\begin{tabular}{c}
$\mathbb{H}_{\alpha\beta}$\tabularnewline
{\footnotesize{}{}(gravitomag. tidal}\tabularnewline
\raisebox{0.5ex}{{\footnotesize{}{}tensor $\sim\partial_{i}\partial_{j}\mathcal{A}_{k}$)}}\tabularnewline
\end{tabular}} & \raisebox{1ex}{%
\begin{tabular}{l}
\raisebox{2.5ex}{}$\bullet${\footnotesize{} Force on gyroscope}\tabularnewline
\raisebox{0,5ex}{~~~{\small{}{}}{\footnotesize{}${\displaystyle DP^{\alpha}/d\tau=-\mathbb{H}^{\beta\alpha}S_{\beta}}$}}\tabularnewline
\end{tabular}} & %
\begin{tabular}{c}
{\Large{}$\times$}\tabularnewline
\tabularnewline
\end{tabular} & %
\begin{tabular}{c}
{\Large{}$\checkmark$}\tabularnewline
\tabularnewline
\end{tabular}\tabularnewline
\hline 
\end{tabular}\caption{\label{tab:Levels}Gravitomagnetic effects present in the Weyl class
Lewis metric, and in the Kerr spacetime, as measured in star-fixed
reference frames (\textquotedblleft canonical\textquotedblright{} and
Boyer-Lindquist coordinate systems, respectively), split by levels
of gravitomagnetism, corresponding to orders of differentiation of
$\bm{\mathcal{A}}$.}
\end{table*}

Even in what pertains to the first level of gravitomagnetism (governed
by $\bm{\mathcal{A}}$), present in both, they substantially differ.
The fact that $\mathbf{d}\bm{\mathcal{A}}=0$ in the Lewis-Weyl metric
means that a Sagnac effect \eqref{eq:SagnacDiffForm} arises only
on loops enclosing the cylinder (as discussed in Sec. \ref{subsec:Physical-properties;-gravitomagn}),
and is independent of the shape of the loop; and similarly for the
synchronization of clocks: the laboratory observers are able to synchronize
their clocks along spatially closed loops that do not enclose the
cylinder {[}in other words, closed in spacetime synchronization curves
exist along the helicoid of Fig. \ref{fig:Hypersurfaces} (b){]};
it is only on loops around the cylinder that a synchronization gap
\eqref{eq:syncgap} arises, see Fig. \ref{fig:Hypersurfaces} (b).
In the Kerr spacetime, by contrast, since $\mathbf{d}\bm{\mathcal{A}}\ne0$,
the Sagnac effect depends on the shape of the loop, and is generically
non-zero (regardless of the loop enclosing or not the axis). The laboratory
observers are likewise unable to synchronize their clocks around generic
closed loops%
.

Another interesting contrast is in the gravitomagnetic clock effect
\eqref{eq:GMClock}. Around the spinning cylinder \eqref{eq:MetricKomar},
since $\vec{H}=0$, it reduces to the Sagnac time-delay \eqref{eq:DtCanonical},
and thus the co-rotating geodesic has a shorter period. In the case
of the Kerr spacetime, by contrast, the term $\Delta t_{H}$ of \eqref{eq:GMClock}
is not zero,
\[
\Delta t_{{\rm geo}}=\Delta t_{{\rm S}}+\Delta t_{H}=4\pi\frac{J}{M};\qquad\Delta t_{{\rm S}}=-\frac{8\pi J}{r-2M}\,(<0);\qquad\Delta t_{H}=\frac{4\pi Jr}{M(r-2M)}\,(>0),
\]
and is actually dominant \cite{PaperDragging}, so it is the other
way around: the co-rotating orbit has a longer period, $\Delta t_{{\rm geo}}=4\pi J/M>0$.
The physical interpretation of $\Delta t_{H}>0$ is that the gravitomagnetic
force $\gamma\vec{U}\times\vec{H}$ in Eq. \eqref{eq:QMGeo} is repulsive
(attractive) for co- (counter-) rotating geodesics, see Fig. 1(b)
of Ref. \cite{PaperDragging}.

\bibliographystyle{utphys}
\bibliography{Ref}

\end{document}